# Plaxis Simulation of Static Pile Tests and Determination of Reaction Piles Influence

Serhii Lozovyi[1], Evhen Zahoruiko[1]

[1]Department of Geoconstruction and Mining Technologies, National Technical University of Ukraine "Kyiv Polytechnic Institute", Borshhaghivsjka str.115/3, Kyiv 03056, Ukraine

**Abstract:** Finite element simulations of four pile static tests were performed using Plaxis 3D. In addition, calculations of pile settlements according to Ukrainian and USSR building codes were performed. These results compared to full-scale pile tests. In order to determine the influence of reaction piles on the test pile response in a static load test were performed simulations with group of reaction piles around tested pile and applied respective negative loads. Plaxis and in situ measured load-displacement curves showed good correlation. Recommendations for Plaxis modeling were given.

**The problem and its connection with scientific and practical tasks.** Before the construction of buildings and structures, where piles are used as foundation, it is required to carry out at least two in-situ static pile tests according to Ukrainian building code. Existing methods of calculating the bearing capacity and settlements of pile foundations, required in building code is rather cumbersome and take a lot of time on their conduct. Simulation of pile foundations testing using computer program Plaxis 3D Foundation reduces time for calculation of pile settlement. This method allow to simulate pile groups with reaction piles and determine their optimal length and diameter, to simulate pile static tests in different parts of the construction site in a variety of soil conditions in order to reduce time and money spent on arranging pile groups and testing.

**Analysis of researches and publications.** In recent years, methods of calculation, design and construction of pile foundations have reached substantial progress. Scientists in collaboration with specialists in design and construction organizations have generalized accumulated experience of pile installation in different soil conditions, have conducted a large amount of experimental and theoretical studies, and have developed new or improved existing methods of calculation and design of pile foundations. One of the big contributions in Ukraine became the new DBN (Ukrainian State building code) V.2.1-10-2009 "Foundations of buildings" with change 1. Substantial contribution to the design and calculation of pile foundations do foreign scientists, including the creation of the program Plaxis 3D Foundation and publications in the exploration of its possibilities [1-3].

**Research task.** The aim of this work is to study new methods of pile foundations modeling using Plaxis 3D Foundation, determination of pile settlements, reaction piles influence on test pile, and comparison of results with the methods of calculating the bearing capacity and pile settlement in accordance with DBN V.2.1-10-2009 "Foundations of buildings" and SNIP (USSR State building code) 2.02.03-85 "Pile foundations"

**Main part and results.** As initial data for the study was taken results of four static pile tests. Initial data included: drawings and schemes of test construction, all necessary geological data, complete test reports with load-displacement graphs. Short description for each testing is given below:

- Root piles #1 and #2, Ø620 mm, length - 23.5 m, were placed during the construction of apartment buildings in the residential area of North Osokorky, Darnitskiy district, Kyiv, Ukraine. Tests were carried out in November 2010.
- Root pile #3, Ø620 mm, length - 9 m, was placed during the construction of the parking lot at SE "International airport "Boryspil", Boryspil, Ukraine. Test was carried out in May 2011.
- Steel tube pile #4, Ø1000 mm, length - 25 m, was placed on construction of the new Chernobyl Nuclear Power Plant sarcophagus "Shelter Object", Chernobyl, Ukraine. Test - June 2010.

Procedure of tests performing presented on Figures 1-2. Physical and mechanical properties of soils and layout of piles in the groups are presented in Tables 1-4 and Figure 3, respectively.

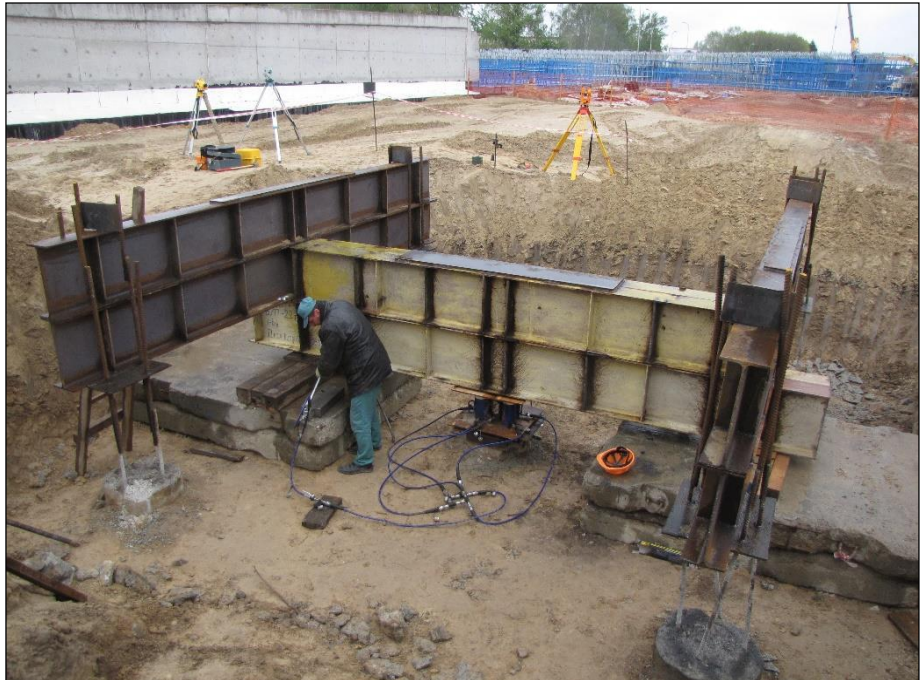

**Figure 1.** Boryspil Pile #3 test view

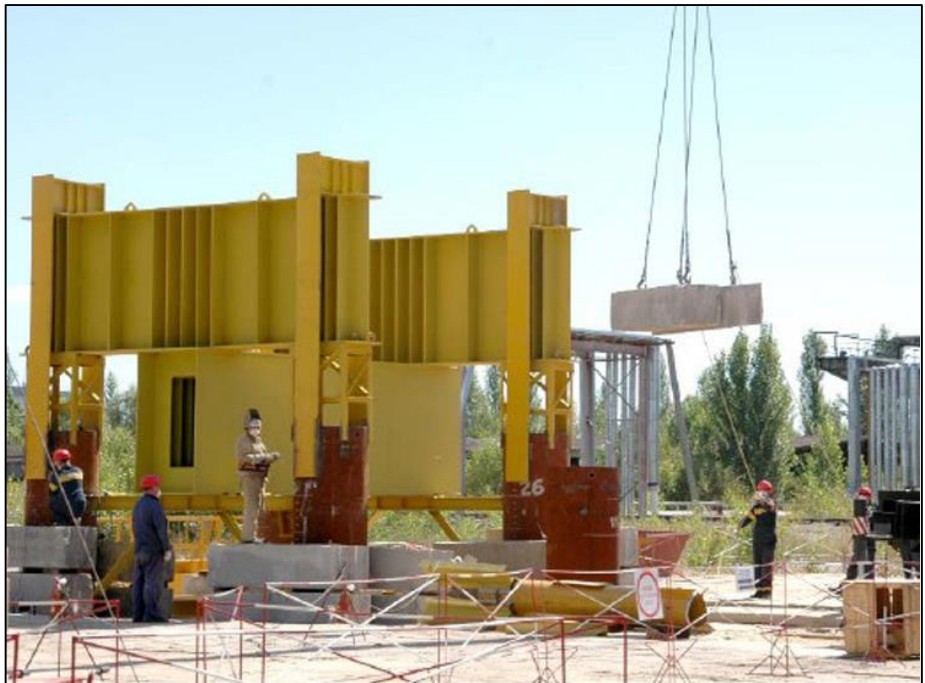

**Figure 2.** Chernobyl Pile #4 test view

**Table 1.** Physical and mechanical properties of soils for pile #1.

| layer # | Soil | Density $\gamma$ [kN/m³] | Permeability $k$ [m/day] | Poisson's ratio $\nu$ | Young's modulus $E$ [kPa] | Cohesion $C$ [kPa] | Friction angle $\phi$ [°] |
|---|---|---|---|---|---|---|---|
| 1 | Reclaimed soil | 17,6 | 10 | 0,3 | 16000 | 1 | 30 |
| 2 | Fine sand | 18,8 | 5 | 0,3 | 34000 | 1 | 31 |
| 3 | Medium sand | 19,5 | 6 | 0,3 | 55000 | 1 | 34 |
| 4 | Sandy loam | 17,5 | 0,5 | 0,31 | 8000 | 7 | 19 |
| 5 | Loam | 19,5 | 0,05 | 0,35 | 9000 | 13 | 17 |
| 6 | Loam with organic additive | 15,5 | 0,05 | 0,35 | 6000 | 15 | 17 |

**Table 2.** Physical and mechanical properties of soils for pile #2.

| layer # | Soil | Density $\gamma$ [kN/m$^3$] | Permeability k [m/day] | Poisson's ratio $\nu$ | Young's modulus E [kPa] | Cohesion C [kPa] | Friction angle $\phi$ [°] |
|---|---|---|---|---|---|---|---|
| 1 | Reclaimed soil | 17,6 | 10 | 0,3 | 16000 | 1 | 30 |
| 2 | Fine sand | 18,8 | 5 | 0,3 | 30000 | 0,1 | 31 |
| 3 | Medium sand | 19,5 | 6 | 0,3 | 45000 | 0,1 | 31 |
| 4 | Sandy loam | 17,5 | 0,5 | 0,31 | 8000 | 7 | 19 |
| 5 | Loam | 19,5 | 0,05 | 0,35 | 9000 | 13 | 17 |
| 6 | Loam with organic matter | 15,5 | 0,05 | 0,35 | 6000 | 15 | 17 |

**Table 3.** Physical and mechanical properties of soils for pile #3.

| layer # | Soil | Density $\gamma$ [kN/m$^3$] | Permeability k [m/day] | Poisson's ratio $\nu$ | Young's modulus E [kPa] | Cohesion C [kPa] | Friction angle $\phi$ [°] |
|---|---|---|---|---|---|---|---|
| 2 | Fine sand | 18,5 | 5 | 0,3 | 30000 | 1 | 32 |
| 3 | Loessial sandy loam | 16,3 | 0,5 | 0,31 | 17000 | 30 | 25 |
| 4 | Sandy loam | 16,5 | 0,5 | 0,31 | 20000 | 18 | 25 |
| 6 | Fine sand | 19 | 5 | 0,3 | 40000 | 1 | 32 |
| 8 | Loam | 18,5 | 0,05 | 0,37 | 22000 | 20 | 18 |

**Table 4.** Physical and mechanical properties of soils for pile #4.

| layer # | Soil | Density $\gamma$ [kN/m$^3$] | Permeability k [m/day] | Poisson's ratio $\nu$ | Young's modulus E [kPa] | Cohesion C [kPa] | Friction angle $\phi$ [°] |
|---|---|---|---|---|---|---|---|
| 1 | Made ground | 18 | 10 | 0,3 | 20000 | 5 | 30 |
| 2 | River sand | 18 | 6 | 0,3 | 35000 | 5 | 33 |
| 3 | River sand | 18 | 6 | 0,3 | 25000 | 5 | 33 |
| 4 | River sand (oozy) | 18 | 4 | 0,3 | 5000 | 18 | 22 |
| 5 | Fine sand | 17 | 5 | 0,3 | 45000 | 3 | 22 |
| 6 | Oozy sand | 17 | 4 | 0,3 | 8000 | 4 | 32 |
| 7 | Medium sand | 17 | 5 | 0,3 | 13000 | 3 | 35 |
| 8, 9 | Fine and medium sand | 17 | 6 | 0,3 | 13000 | 3 | 33 |
| 10 | Clay marl | 15,3 | 0,005 | 0,35 | 80000 | 38 | 23 |

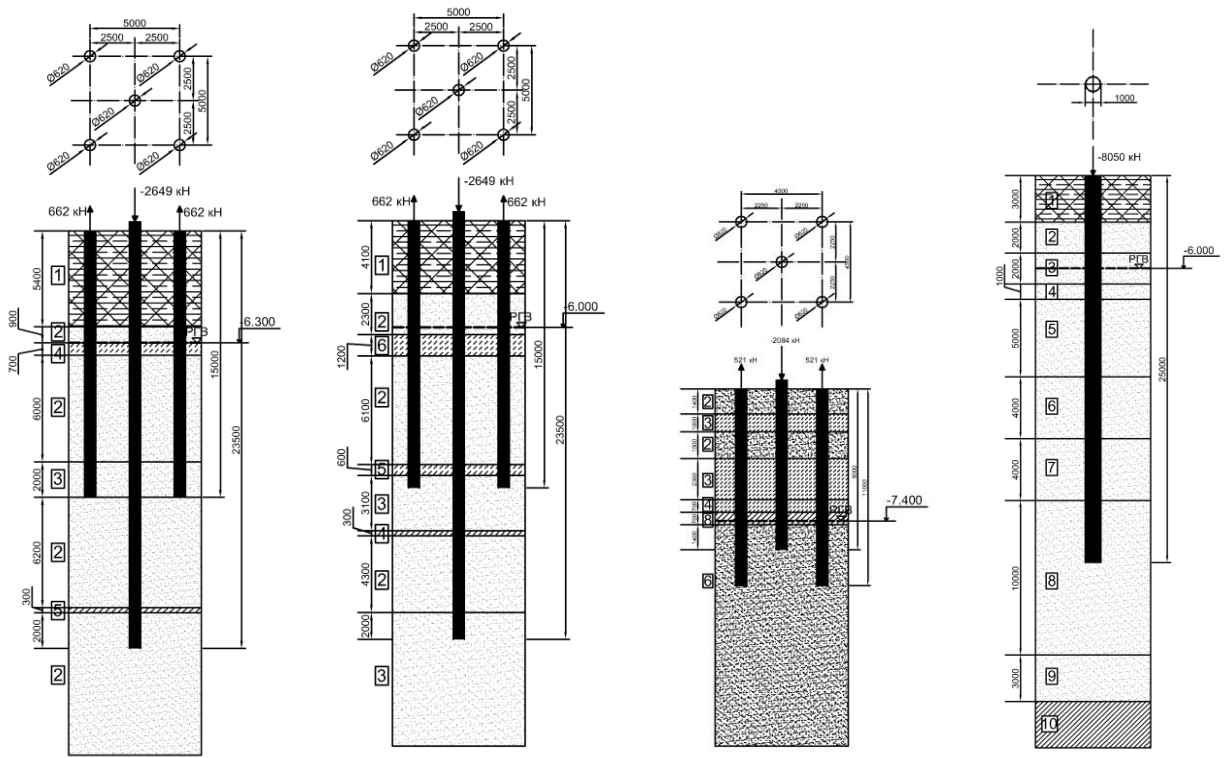

**Figure 3.** Geological section with the schemes of placing piles in groups.

Were calculated bearing capacity of three root piles (№1, 2, 3) according to DBN V.2.1-10-2009 and SNIP 2.02.03-85, also were calculated pile settlements from static loads according to the layer-by-layer summing method from DBN V.2.1-10-2009.

The result of pile foundations testing simulations using computer program Plaxis 3D Foundation are load-displacement plots, these results are compared with the results obtained during the full-scale pile static tests on the graphs - Fig. 4-7. Also shown the correlation coefficients separately for pile loading and unloading. The red curves show displacements in simulation using the finite element program Plaxis. The blue curves show displacements determined in full-scale static tests.

To simulate soils in Plaxis was used elastic-plastic Mohr-Coulomb model.

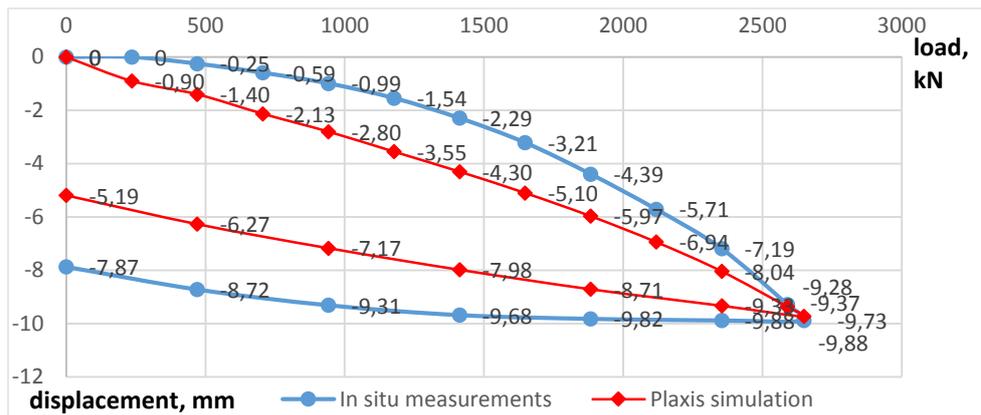

**Figure 4.** Load-displacement curves for pile #1.

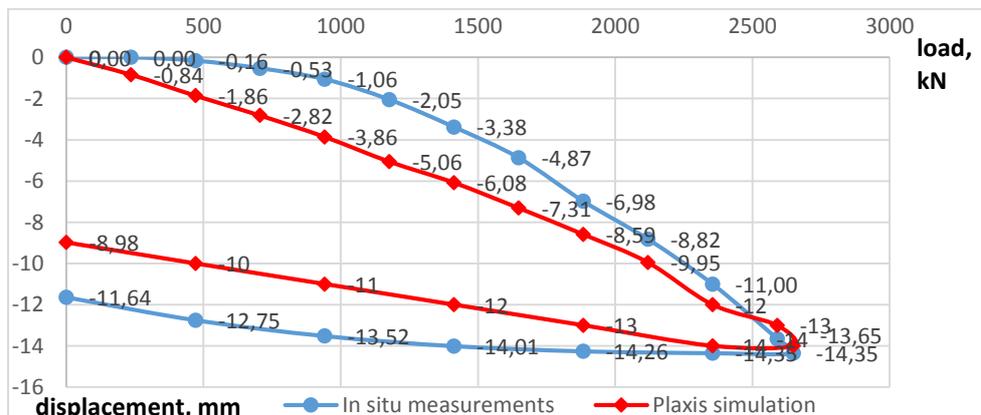

**Figure 5.** Load-displacement curves for pile #2.

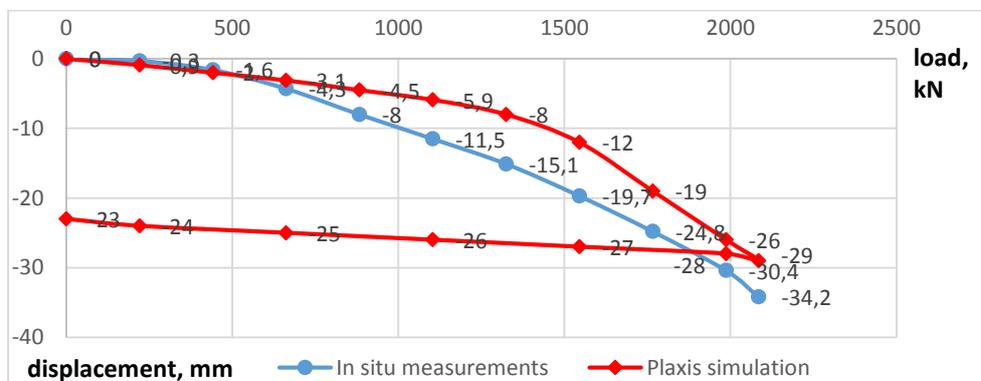

**Figure 6.** Load-displacement curves for pile #3.

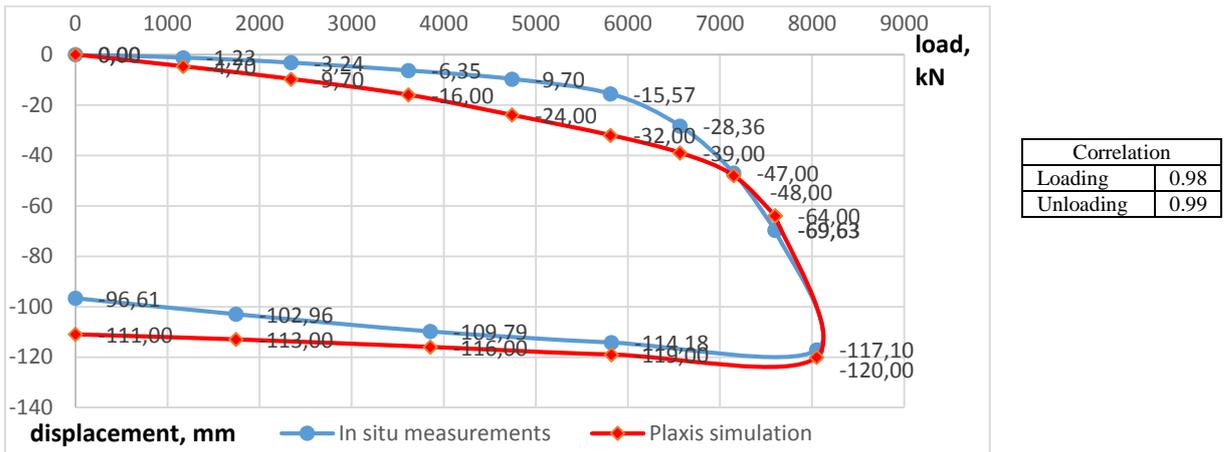

**Figure 7.** Load-displacement curves for pile #4.

Comparison of results full-scale tests, calculated bearing capacities according to DBN V.2.1-10-2009 and SNIP 2.02.03-85, and Plaxis 3D Foundation program are shown in Table 5. Also compared pile displacements from in situ results, Plaxis, and DBN V.2.1-10-2009.

**Table 5.** Comparison of results.

| Pile # | Pile bearing capacity according to DBN V.2.1-10-2009, kN | Pile bearing capacity according to SNIP 2.02.03-85, kN | Pile bearing capacity according to full-scale tests, kN | Maximum pile displacement during full-scale tests, mm | Pile displacement calculated according to DBN V.2.1-10-2009, mm | Maximum pile displacement during Plaxis simulation tests, mm |
|---|---|---|---|---|---|---|
| 1 | 2725 | 2089 | 2207,5 | -9,88 | -8,03 | -9,73 |
| 2 | 2808,5 | 2146 | 2207,5 | -14,35 | -11,06 | -13,65 |
| 3 | 1297 | 1056 | 1737 | -34,20 | -32,62 | -30,40 |
| 4 | | | | -117,10 | | -120,00 |

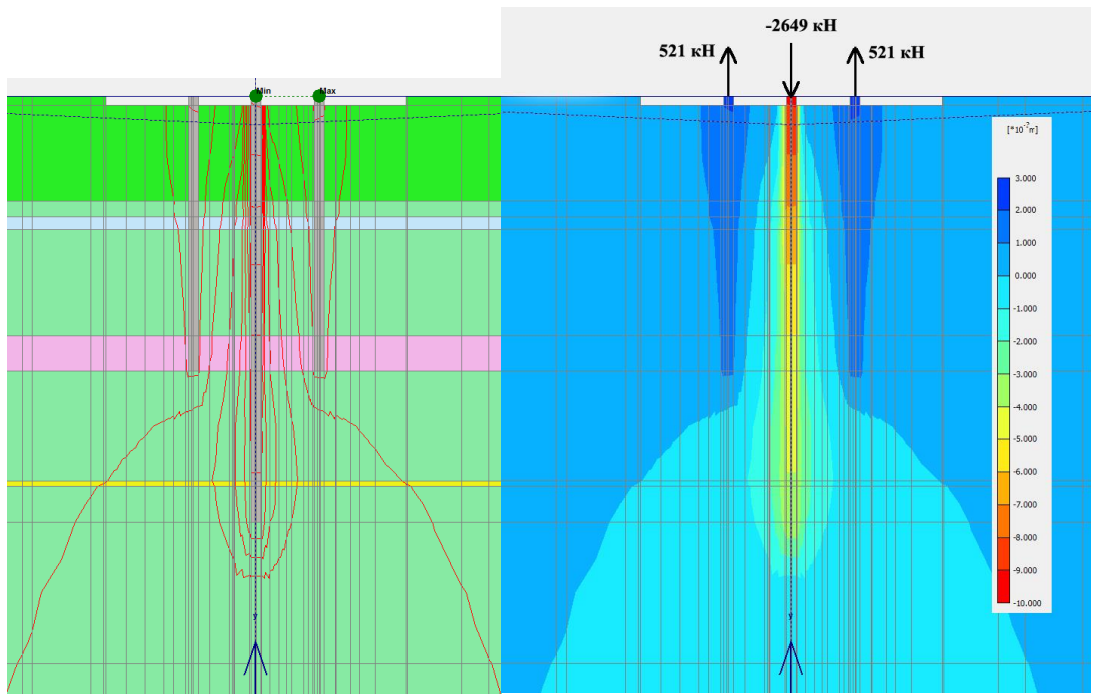

**Figure 8.** Distribution of deformations in soil at pile #1 static tests simulation.

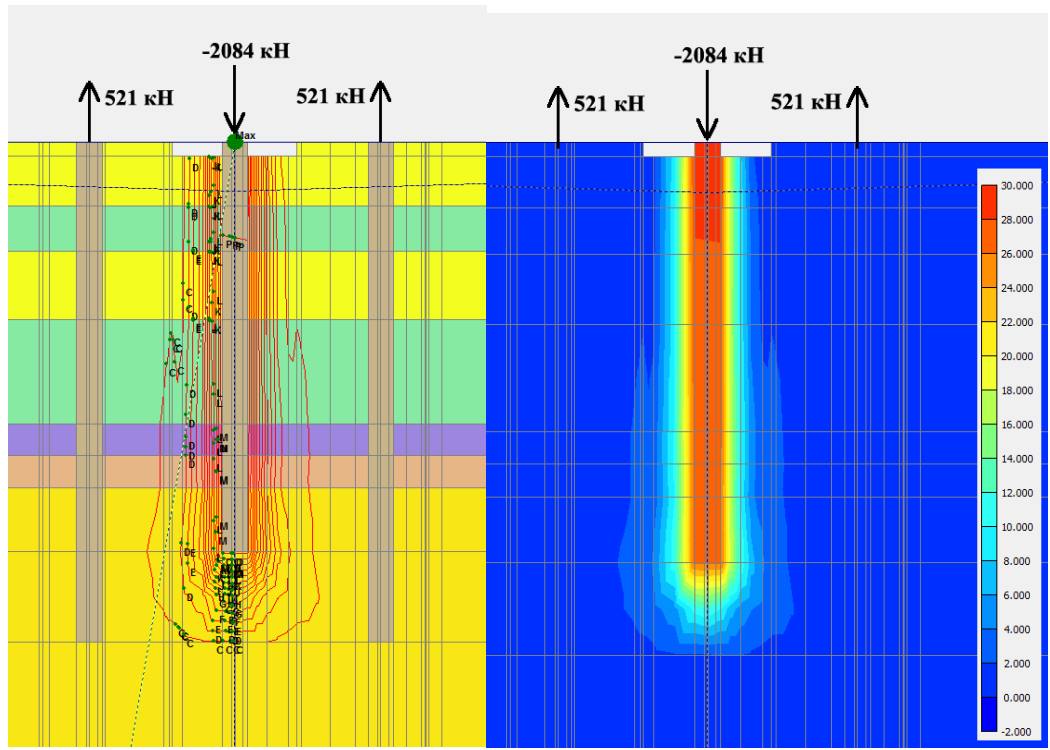

**Figure 9.** Distribution of deformations in soil at pile #3 static tests simulation.

Simulation modeling of four piles static tests performed in the program Plaxis 3D Foundation resulted in the load-displacement graphs. Curves obtained in simulation are compared with graphs of full-scale static tests. The average correlation coefficient is 0.98, and the average difference between maximum values of settlement is 5%, indicating a high reliability of the data obtained in the simulation.

The accuracy of calculation using program Plaxis 3D is better than layer-by-layer summing method according to DBN V.2.1-10-2009. The average deviation of DBN results from full-scale tests is 15.7%.

The deformations distribution were almost the same for pile #1 and #2, so were presented only for pile #1. As seen from fig. 8 reaction piles are shorter then main pile and we can observe their displacement in upward direction, which is unacceptable. The correct design of reaction piles on fig.9 for pile #3 shows that there is no displacement for these piles, only the main pile goes down.

**Conclusions.** The obtained results allow to recommend Plaxis 3D Foundation in simulation static pile tests in different soil conditions of the construction site to reduce costs for additional field tests, calculation of bearing capacity and pile settlement calculations, plotting load-displacement curves.

In addition, simulation-modeling program Plaxis 3D Foundation allows to reduce time for design and calculation of pile foundations, increase efficiency, reliability, and informativeness of calculations, reduce the cost of performing additional field tests, and to determine the optimal size of the main and reaction piles for the tests. Plaxis makes possible to conduct simulation modeling of required number of pile static load tests in different parts of the construction site and in different soil conditions. The results of Plaxis simulation should be considered valid only when performed two required full-scale tests and they match the simulation results.